# ENTERPRISE MODEL VERIFICATION AND VALIDATION:
## AN APPROACH

**V.Chapurlat\*, B.Kamsu-Foguem\*, F.Prunet\*\***

\* LGI2P - Laboratoire de Génie Informatique et d'Ingénierie de Production
site EERIE de l'Ecole des Mines d'Alès - Parc Scientifique Georges Besse - F30035  Nîmes cedex 1
Tel : (+33) 466 387 065 - Fax : (+33) 466 387 074
Mél : Vincent.Chapurlat@.ema.fr
\*\* LIRMM - Laboratoire d'Informatique, de Robotique et de Microélectronique de Montpellier
161, rue ADA, 34192 Montpellier cedex 5
Mél : prunet@lirmm

**Abstract**: This article presents a Verification and Validation approach which is used here in order to complete the classical tool box the industrial user may utilize in Enterprise Modeling and Integration domain. This approach, which has been defined independently from any application domain is based on several formal concepts and tools presented in this paper. These concepts are property concepts, property reference matrix, properties graphs, enterprise modeling domain ontology, conceptual graphs and formal reasoning mechanisms.

## 1. Introduction

Enterprise Modeling and Integration is now considered as an important research and application domain by industrial users. They dispose of several approaches (AMICE 1993, Loucopoulos 1995, Vernadat 1997, Crestani 1997, Bernus 1998, Sheer 1998, Menzel 1998, GERAM 1999, Tissot 1999, Vernadat 1999, Chen 2001, Chorafas 2002, Revelle 2002, Chen 2002) consisting of modeling languages, norms, architecture reference models and derivative tools allowing them to describe their processes, their information systems, their knowledge and know-how, to better understand and to test the organization behavior by using for example simulation mechanisms, to communicate more efficiently in the enterprise, to exchange information and data without loss of sense with each other partners, to decide several system and organization improvements and so on.

Nevertheless, the user can doubt of the amount of confidence he can have in the different built models such as processes models or human resources ability models for example. Indeed, before considering a model is well suited to be use, this one must be bound by some phases during which it has to be verified (« is the model correctly built ? ») and to be validated (« is the model corresponds accurately to the reality ? is it the good one taking into account the needs and the context ? ») summarized Figure 1. Results of this global step of verification / validation (V&V) depends on two causes. First, the previous modeling step induces several problems on account of loss quality of the model which have to be taken into account during verification and validation:

- The user may not be clear in one's mind and his point of view may evolve during modeling process,

- Modeling hypothesis (temporal, behavioral rules and so on) limit model's expressiveness and accuracy. They are caused by system typology, system complexity level, by modeling concepts or formalisms which may be restricted to particular kinds of systems,

- The verification and validation need to highlight a user's defined analysis perspective (performance, temporal independence, functionalities, reliability and so on). The model may be unsuitable to respond correctly to some of the possible analysis perspectives,

- It stay often difficult to take into account the modeled system environment, its own dynamics, the different interaction between components and possible unforeseen events which may be the origin of some emerging and unexpected phenomena,

- Furthermore, the lack of knowledge and the oversight about particular characteristics, data and information

from the system itself or from its environment must limit again the model V&V process.

What comes out of these causes is that a part of the knowledge needed for assuming the quality and the relevance of a model (from a static point of view in a first way then taking into account dynamic evolution rules) remains informal and misunderstands during modeling process.

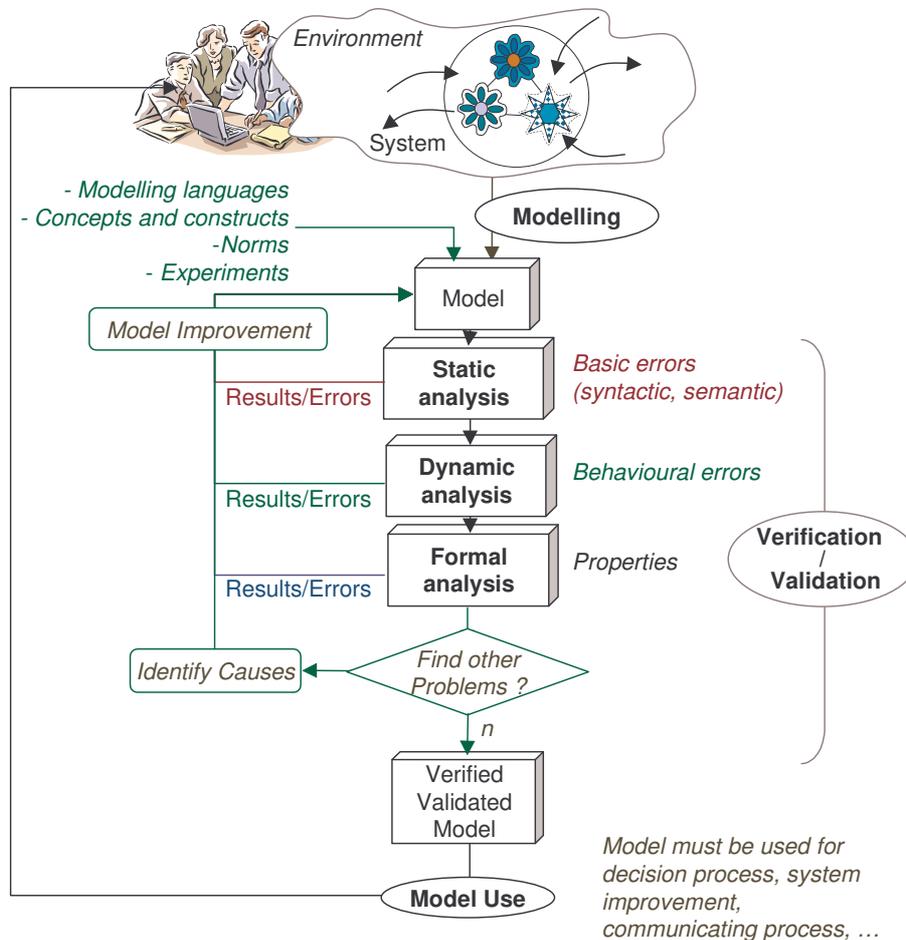

**Figure 1 :** A model life cycle vision highlighting verification and validation phases

This lack of knowledge may be justified by taking into account the user's behavior during modeling step. He can objectively rule out some information, data and events because of their nature, the time needed to describe them, their relevance considering his own point of view and so on. He can also forgot them because of its limited vision of the system or the different situations in which this system must evolve. It may be then interesting to develop some mechanisms allowing us to show the user the relevance of some of these information and data because of their possible interaction with the information and data gathered in the model it self in order to complete and/or to improve this model.

Second, each V&V phases shown Figure 1 set to work concepts and tools allowing syntactic checking (not considered here), semantic checking and behavioral analysis which use mainly simulation, emulation, human expertise of model execution results in particular in enterprise modeling and integration domain. The analysis results must be altogether questionable considering several points such as formalization and detail levels of the model for example. The way consisting on developing some other kind of tools allowing us to employ formal properties proof mechanisms seems to be interesting to use in this domain such as proposed in (Van der Aalst 2000, Lemboley 2001, Covès 2000).

So, the research results presented in this paper intent to cover the V&V step by using a formal verification and validation approach allowing to improve user's knowledge about its model and to manage it in order to establish the relevance and the suitability of a model. The modeling and model/system improvements steps are not considered in this article.

## 2. Verification / Validation step approach

The proposed approach is based on several concepts detailed in the following. These concepts have been defined independently from a given application domain and, as we intend to show in the example at the end of this article, they have been specialized and putted into concrete form in Enterprise modeling domain. These elements shown in Figure 2 are:

- The property model is the base of a formal modeling language allowing the user to specify a particular knowledge he wants to prove on the model in order to verify and to validate it.

- A domain ontology (Uschold 1996, Fox 1994) describes a given whole application domain by defining the different concepts and relations between these concepts which are necessary.

- This ontology is then translated by using formal rules into two lattices needed for the V&V chosen tool based on conceptual graphs by splitting up concepts and relations.

- In same time, this ontology allows us to define a set of generic properties associated to the chosen domain and are gathered into a data base named Reference matrix. These properties cannot be proved directly on the model. Their goal is to guide the user, to remember him some crucial information which may be forgotten or misunderstand and to help him to specify completely the relevant and useful properties the model must respect.

- These generic properties will be interpreted, instantiated or translated respecting the model and will be gathered in the Property Graph. This one represents all the properties which have to be proven on the model.

- Some reasoning mechanisms have been developed taking into account Conceptual Graphs theory in order to prove the proposed properties on the model and to make emerging when it is possible some additional knowledge.

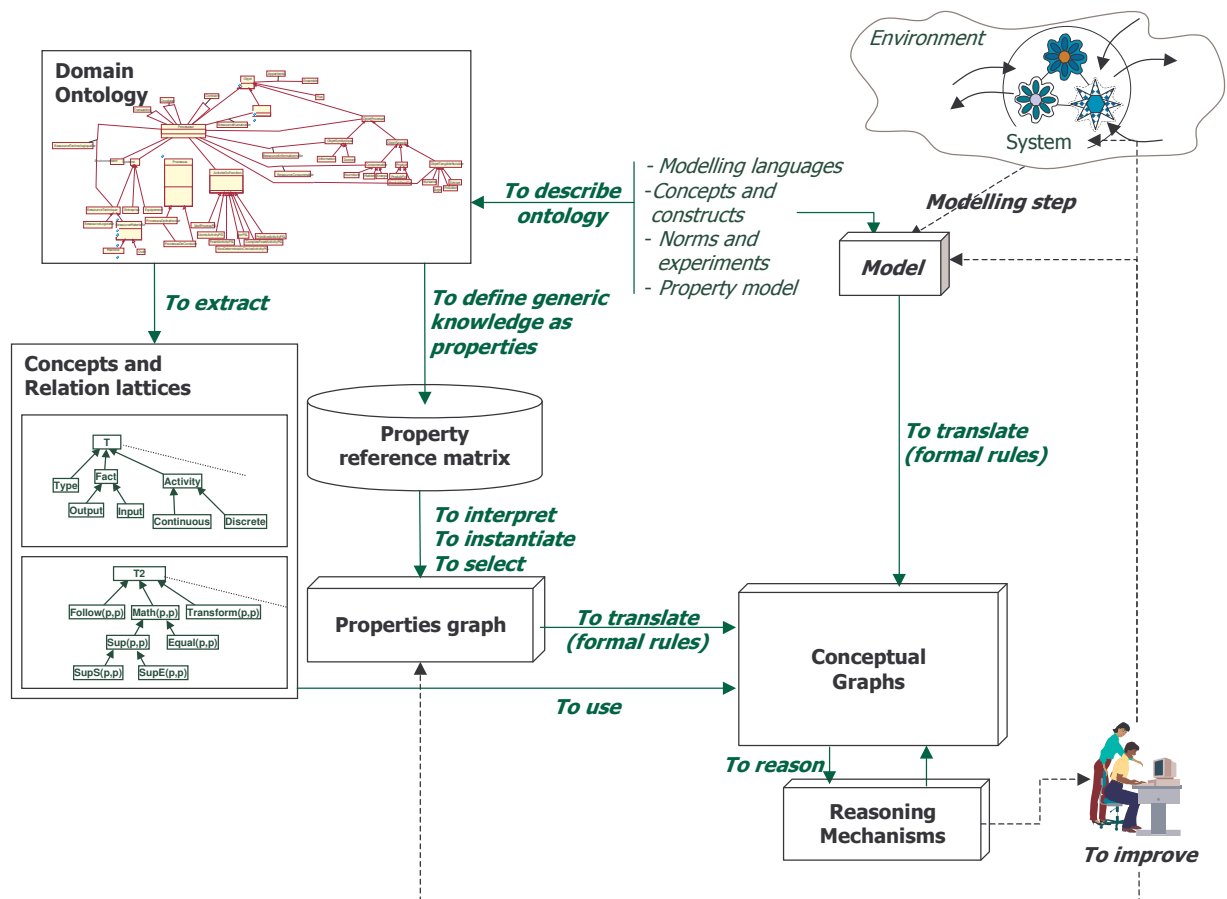

**Figure 2 :** Global approach and components

## 3. Property model

A property may be defined in the literature with several points of view (Feliot 2000, Meinadier, 1998, Thome, 1993, Manna 1982, Henzinger 1994, Manna 1982, Manna 1990). We will consider in the following the definition given by (Lamine 2001): *a property translates an expectation, a requirement (behavioral, functional, structural or organic, dependent or not of time) or an objective (performance, safety, reliability and so on) which have to be respected, strictly or with a reliable level being enough by a model*.

### 3.1. Informal model property

The proposed model property we will use in the following is based on a *causal relation* named R between two sets named C (the set of *causes*) and E (the set of *effects*). Each of these sets is composed by a collection of objects called *facts* issued directly from the model. As proposed by (Pearl 2000, Pearl 1999, Sowa 2000), the causal relation R indicates how occurrence of elements of E depends on the occurrence or on the interaction between elements of C. The property may be verified by computing in which cases a *condition* on the causes and the *result* on the effects are true by respecting temporal hypothesis and causal relation typology between *causes* and effects. Before presenting the formal model, it is necessary to define two particular concepts named *facts* and *granularity*.

### 3.2. Facts

Facts come from several origins:

- From the *modeling language concepts* (entities of modeling such as process, activities or resources) and *relations* between these entities (such as the link between activities and resources for example) as the modeling language evolution laws and rules. They are called *handle functions*. For example, the stateOf(A,t) handling function allows to define what is the state of the Activity A in time t.

- From the *modeled system context* and *application domain*. This needs to define a domain ontology allowing us to define the existing concepts, relations and possible situations in the domain which are not completely or already described into the model. All this added knowledge is then translated into new facts allowing us to merge it with the information coming from the model. They are called *modeling variables* and *modeling parameters*. It may represent for example the external temperature of the modeled system which is an input of the model but which has been forgotten during modeling step. It may also represent some external events from the environment or internal events such as data evolution.

- From the *model* itself. This needs to extract all the information contained in the model as new *modeling variables* and *modeling parameters*. It may be for example a modeling parameter containing the maximum level of water in a tank or a modeling variable containing a computerized data such as production rates or the input model vector state at each moment.

- From other existing properties in which the user can trust at a given moment: these properties are then considered by new facts as *properties*. This may be for example a property specifying what is the speed limitation taking into account some constraints at each moment.

At last, fact may be valued quantitatively or qualitatively (a property is true, a data is set to '30' or to 'good'). Most formally, the set of facts *F* is defined as follows:

$$F = MV \cup MP \cup HF \cup P$$

where *MV*, *MP*, *PR* and *P* gather four kinds of facts :

- *MV* is the set of facts named modeling variables : each ones evolves within the modeled system:

$$MV = \{ var / var = < name^1, type^2, value, Def^3 > \}$$

- *MP* is the set of facts named modeling parameters : they described data which have constant values:

$$MP = \{ par / par = <name, type, value> \}$$

- *HF* is the set of facts named *handle functions*. They allow to manipulate all the facts and then to characterize

---

[1] name is a tag defining each fact on an unique manner
2 type is the fact's type: ℝ, ℂ, Boolean, Character or structured type
[3] def is the fact's definition domain (Def ⊂ Type)

the model or the environment dynamic and structure. For example, if the model is a transition model such as a Petri Net, it exists function allowing to describe net structure (before(place), follow(place), weight(arc), tempo(transition) and so on), to describe marking evolution (mark(net,t), fire(transition,t) and so on):

$$HF = \{hf \ / \ hf = \ < name, \ paramaters[4], \ type \ >\}$$

- *P* is the set of user-defined properties

### 3.3. Granularity

The property concerns a given and unique model. It is an interpretation of some user's requests and must be then proven only on this model However, establishing a complex system model need often to define a hierarchy of description by using decomposition or substitution rules. User may then define system's model(s) then sub(system(s) model(s) and so on in order to manage more efficiently the system complexity. In an other hand, in user's minds there exist some properties which may be assignable to only one detail level of the global system description but the interaction between these properties of a given level could be at the origin of the appearance (that is to say the emergence) of some new properties of a lower level. The *granularity* G is then defined in order either to allow the user to respect the different detail levels which are necessary to manage the system description, whether to allow this user to choose a particular sets of own detail levels. A granularity G is then the resulting set of chosen detail levels, called *degrees.*

The example presented Figure 3 shows a granularity allowing us to define four levels of decision from strategic one to execution one. Strategic level is the needed set of activities and processes allowing to define or to better understand enterprise finality and its objectives. Tactic level has to take into account some strategic decision and to define piloting orders and planning allowing the organization to reach these objectives. Operational and execution levels have then to schedule different production processes, to define stocks size, to manage production resources and so on. A temporal dimension may be associated to each level in order to help the user when we will define a property of a given level.

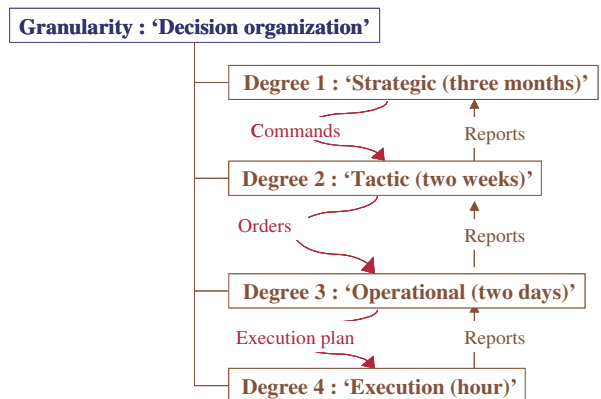

**Figure 3 :** Granularity and degrees

At the Execution level, there are lot of exchanges of flows and several humans skills needed during the production process execution based on the resources cooperation. For example, in case of a long enough mechanical failure of a given machine (a given property is then not verified) it will be then possible that operational level intents to re define all scheduling orders which cannot be applied at the execution level.

### 3.4. Formal Property Model

A property *Pr* is defined by a 5-tuple:

$$Pr = \ < name, \ C_p \ , \ R_p \ , \ E_p \ , \ D_p >$$

Where :

- $C_p = \{$ cause / cause $\in F\}$ / card$(C_p) \geq 0$ (set of causes may be empty)

- $E_p = \{$ effect / effect $\in F$ $\}$ / card$(E_p) > 0$ (a tangible effect must always exist) and $C_p \cap E_p = \varnothing$

- $D_p = \ < Type \ , \ G >$ is the degree of *Pr* in a given granularity G

---

[4] parameters is a set of facts

- $R_p$ is the relation defining the causal link between causes and effects. Type of $R_p$ may be of:

    - **Logical**: it describes implication and equivalence (a reciprocity between cause and effect) relations.

    - **Temporal** : it describes temporal links such as antecedence in which the cause must be prior to, or at least simultaneous with the effects.

    - **Influence** : *the knowledge about some cause modifies the opinion about the verification of the effect* (Pearl 2000). It allows then to describe how causes and effects must be linked respecting some particular events or situation. A sense of variation is associated to each influence relation. It can be interpreted as beneficial or at the opposite harmful influence on the effect.

    - **Emergence** : Each modeled system can be described by some characteristics which are not directly deductible from the own characteristics of its components but which result from relations between these components. The explanation of this kind of property needs then to take into account all the interactions and the feedback which connect the referent with its environment or with its context.

The relation $R_p$ is defined by the 4-tuple:

$$R_p = < Type , \; \theta_c , \theta_e , d > \text{ where:}$$

- The Boolean functions θc describes in which conditions (by interpretation of causes), the property is verified. θc is defined as follows :

    - θc : C → {true, false} ; If there is an empty cause then θc = true

- The Boolean functions θe describes with which results (by release of effects), the property is verified. θe is defined as follows :

    - θe : E → {true, false}

### 3.5. Short examples

In a manufacturing process, an activity $A_i$ transforms one or several of its inputs in order to furnish one or several outputs taking into account constraints, rules and mechanisms as shown in Figure 4.

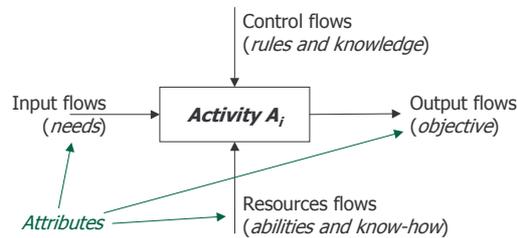

**Figure 4 :** Activity seen as a transformation scheme

This activity induces then modifications in the time, in the form or in the space of some of these inputs attributes which corresponds then to one several of the outputs attributes (Mayer 1995, Feliot 1997). This implies some relation between the inputs and the outputs. For example, if the output is of energy type, this induces that one of the input may be of material or of energy type in the same time. This kind of property may be wrote as follows:

*[ ∀Activity.OutputInformation / OutputInformation.OperationalDomain = Energy]*

*implication (⇒)*

*[ ∃Activity.InputInformation, (InputInformation.OperationalDomain = Material ∨ InputInformation.OperationalDomain = Energy) ]*

Last, the following property may allow the user to detect an error in the model. The property is the following: 'a *transport link ensures continuity of parts flow in manufacturing system between two sub-systems'*. Figure 5 shows the model which may respect this property. It shows a manufacturing process composed of a drilling activity and a polishing activity. If the drill station is not near the polisher station or if the two functions (to polish and to drill) are not assumed by the same machine, the user forgets some transportation activity and support as shown in Figure 6.

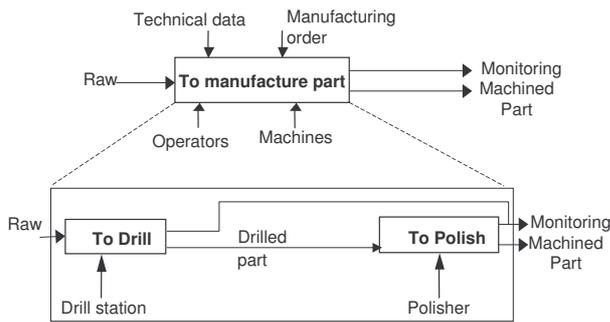

**Figure 5 :** A part of a process model to check

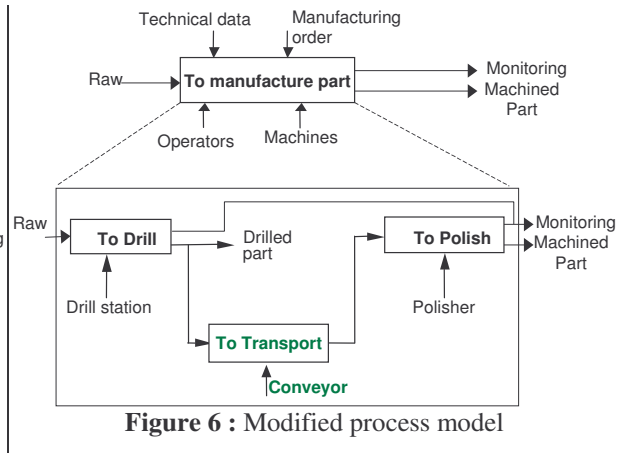

**Figure 6 :** Modified process model

The modeling language for the property specification has been defined and integrated into the LUSP Language support tool (French acronym of Properties Specification Unified Language) used for specifying, managing and formally proving a property (Chapurlat 2000).

# 4. Domain ontology

Before representing knowledge as properties, it is necessary to determine an ontology (Uschold 1997) dedicated to what we need to represent in a given application domain. Thus, the formalism chosen here was adapted to take into account the systemic concepts (Mayer 1995, Le Moigne 1999, CEA 1998, Braesch 1995) and the other concepts dedicated for Enterprise Modeling domain (Schlenoff 1996, Tissot 1999, ISO 2002) by the construction of an object class model by using the Unified Modeling Language (UML). This formalism has been used taking into account its readability, the opportunity to represent concepts and relation and the industrial recognition of this modeling language.

Thus constituted, the obtained set of concepts and relations can represent a common base of a modeling language allowing the user to describe each enterprise organization part. This point allows the authors to link and to justify the presented work concerning V&V approach with the current work in progress on Unified Enterprise Modeling Language (UEML) (Vernadat 2002, Vallespir 2003, see also www.UEML.org) which aims on the definition of a common enterprise modeling language. Figure 7 shows a simplified version of this concept lattice composed by a pre-defined set of concepts. It is divided into four parts: abstract types, behavioral types, modeling types and the entity types.

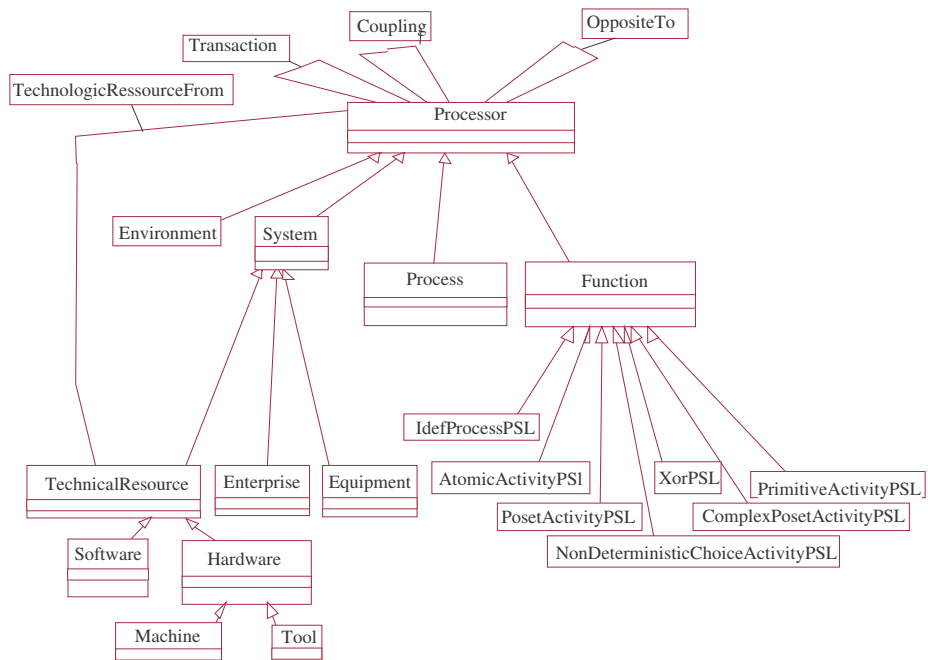

**Figure 7 :** Extract from one of the object diagrams showing the ontology

## 5. Properties reference matrix

As shown before, specifying a property needs to extract all the facts from the model and to add some new facts extracted from its environment by using the domain ontology. However, the specification work remains difficult for a non specialist user taking into account formal concepts such as those presented in (Van Lamsweerde 2000) and the model presented before.

First it is necessary to fix some idea about the goal of the V&V step which has to be done. There exist three possible perspectives of analysis (CEA 1998) which can be used independently from the other ones or which can be combined:

- **Stability** which aims on the ability of the system to maintain its viability all along its life (that is to say to maintain a sufficient relation between its structure and its coherence).

- **Reliability** which aims on the ability of the system to reach its objective (that is to say to ensure its function with a high level of compliance).

- **Integrity** which aims on the ability of the system to stay coherent and to be able to ensure its function.

Second, the goal is then to develop a properties data base called Properties reference matrix (Chapurlat 2002). It gathers usual knowledge describe by using property model considering a given application domain, a given set of modeling language and taking into account a systemic vision of the pointed out system and/or model to analyze. These properties cannot be proved on a given model because of these generic nature. They must be *instantiated* (the user create an occurrence of the generic property by filling in its causes, effects and relation) or to be simply *interpreted* (the user can be inspired by this generic property).

The property reference matrix uses a property typology which has been inspired from a literature and research works analysis such as (Paynter 1961, Lamport 1980, Manna and al. 1990, Manna and al. 1992, Sahraoui, 1994, Feliot 1997, CEA 1998, Lamboley 2001). This typology discerns three kinds of properties :

- **System properties** : These properties express the constraints and the functional or not functional requirements in which each system of a given application domain is (or will be) subjected and its assigned objectives. They are properties of functioning (temporal or not), of security, of volume, describing needed performance (productivity, availability, reliability and so on).

- **Modeling Language properties** : These properties describe the model structure (by defining and representing the modeling language construction rules and the possible existing constructs proposed in the modeling language), the behavioral laws (by defining execution rules effects on the model (Chapurlat 1999)) and the possible properties of liveliness, completeness, coherence, of reinitializing, describing the presence or the absence of parallelism, of synchronization mechanisms, of sequence, of temporary or definitive blocking and so on. For example, considering Petri Nets modeling language, the Model Properties describe structural rules concerning places and transitions placement, marking vector evolution, temporal hypothesis of synchronism or parallelism and so on.

- **Axiomatic properties**: They permit to describe basic knowledge, that is to say a set of information collectively and unanimously recognized and accepted such as laws of nature, norms, standards and also some existing property which is already verified for the user. Thus, they are indisputable and the user may choose them as facts for describing and proving other properties.

## 6. Properties graph

The properties reference matrix help the user to choose and to specify the relevant properties he wants to prove. The results of this specification phase is a graph in which:

- each nodes represents the different sets of facts considered as causes or as effects of a given property,

- each arc represents a typed relation between these causes and these effects .

This graph takes into account simultaneously the knowledge about the model, the modeled system and its environment. All this knowledge is described by using the unique property model (and then the modeling language LUSP). That allows to take into account all this knowledge simultaneously and with the same mechanisms of analysis. In order to structure the user's work a properties graph may be represented as a third axes area in which:

- The first axe called *target* allows to separate into three level of detail what are the different objects (systems or models of these systems) which represents the target of the V&V step or which interact with this target and can influence then the V&V results. These objects are chosen by the user by respecting the selected analysis perspective (stability, integrity, reliability or a combination of them). The central level called referent level highlight the pointed out system which has been modeled and the resulting model representing the V&V goal. The upper level called upper referent allows to describe the pointed system's environment regarding the user's point of view. For example if the user has chosen stability perspective the upper referent may represent either all the encountered process in the company whether the process's customer inputs/outputs. For each referent level it may then possible to define several upper referent. The lower level is composed of other objects (sub systems, sub models) which compose an interacting network either corresponding to the decomposition whether having the same global behavior from the object belonging to the referent level. Each of these three levels is connected with the above respecting rules. These ones depends on decomposition rules imposed by a given modeling language if the referent level object is a model or user decomposition point of view if it is the system.

- The second dimension called *typology* permits to separate system and model properties. For each ones, it is then necessary to clarify the properties which are connected to the structural aspect, to the behavioral aspect or to the functional aspect expected from the object O chosen considering the target axe. Systems properties may be interpreted or instantiated taking into account the properties reference matrix. They represent the current state of the system or the needs at which the system must respond in term or performance and constraints. Model properties are issued from the translation of modeling language properties.

- The third axe is called *time*. Past, present and future of an object O must be taken into account in order to manage the possible evolution of the properties of O. It allows the user to reuse part of existing properties and to complete them during life cycle evolution of the target.

Figure 8 summarizes these three axes and each case may be considered as a property graph. Figure 9 shows an example of these different object. The referent level is then defined by a given process to improve and the user wants first to verify its model. Doing this needs to define what is the environment of this process and what the possible sub systems. In the same idea it may exists some model describing this environment or each of these sub systems. The information gathered into these new models must be then employed for the V&V phase.

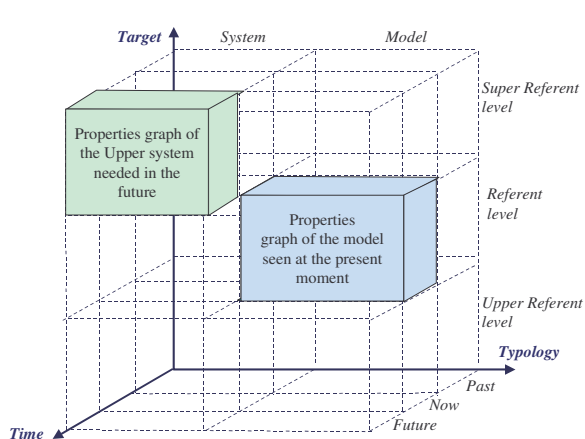

**Figure 8 :** Property graph represented as a third dimensions area

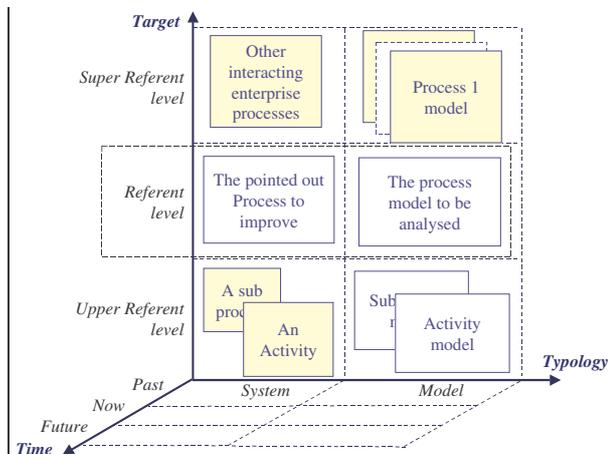

**Figure 9 :** Property graph objects example

## 7. Conceptual graphs

It is necessary to dispose of a set of mechanisms allowing us:

- To manage all the previous highlighted knowledge for V&V steps. As shown before, this knowledge occurs from user's specified properties which are gathered into the properties graph. However additional knowledge depending directly from the system or from its environment but which may be absent and not already defined into the property graph have to be added again.

- To prove formally the proposed properties. This phase may use different kinds of concepts and tools (Rushby 1995). It may be based on some theorem prover such as Zeves (dedicated only for the Z modeling (Barden 1994) language which is more appropriate for proving the structure coherence and adequacy of a given modeling language even a model than properties as proposed below). It may also use model checkers such as PVS (Owre 1999), Stanford Temporal Prover STeP (Bjorner 2001), SMV (see http://www-2.cs.cmu.edu/~modelcheck/smv.html), SPIN (Holzmann 2003) and other. All these tools are formal proof tools based mainly on state diagrams or transition models properties verification and data analysis. Their utilization, very limited in Enterprise Modeling domain, can then make easier proof of behavioral properties if the modeling language uses the same concepts of state, transition and temporal hypothesis. Last, this phase may use simulation or emulation tools allowing the user to execute the model by taking into account some pre-defined scenarios. This kind of solution is now a well known one in the industry but the problem comes first from the scenarios, second from the execution mechanisms used. Scenarios may be subjectively forgotten or wrong defined (for example, some particular and relevant events or system's situations may be ignored). Execution mechanisms i.e. the set of evolution rules are not sufficiently formally defined (Petit 1997, Panetto 2001) in an indisputable way. So, the simulation may give some questionable results.

- To make emerge new knowledge allowing the user to complete again and again its system's representation.

This research work intents to ameliorate and to extent the existing V&V concepts in Enterprise Modeling domain. So, the Conceptual Graphs have been chosen for the followings reasons (Kamsu-Foguem 2003). First, it allows to conceptualize the pre-defined domain ontology as a formal vocabulary. Second, it disposes of a range of formal reasoning mechanisms (rules, projection, first order logic isomorphism and other) permitting to verify facts and to complete (or to make emerging) a new knowledge by combination of existing knowledge.

The goal now consists to define the adapted vocabulary and to translate the model and the properties graph obtained before in order to use the reasoning tools associated to conceptual graphs.

### 7.1. Formalism Presentation

A simple conceptual graph is a finite, connected, directed, and bipartite graph composed of two kinds of nodes called *concepts* and *conceptual relations*. A **concept** is composed of a type and a marker:

$$[<type>: <marker>]$$

where:

- *type* represents the concept typology which is necessary to describe a given domain. They are grouped into a hierarchical structure called concept lattice (see a simplified version of it in).

- *marker* specifies the meaning of a concept by specifying an occurrence of the type of concept. They can be of various natures, in particular individual or generic.

For example [*Process*: *Customer needs definition*] means that *Customer needs definition* is one the numerous processes of the enterprise. Its represents then an occurrence of the *Process* concept into the modeled domain.

A conceptual **relation** binds two or many concepts according to the following diagram:

$$[C1] \rightarrow (relation) \rightarrow [C2] \ (means "C1 interact with C2 by relation")$$

For example, [*Activity*: '*To specify needs list*'] → (*usage*) → [*Actor*: *Specialist*] means that *Specialist* who is an *Actor* uses the *activity* named '*To specif needs list*'. Each relation is characterized by a *signature* which fixes its arity and gives the types of concepts which are in relation.

The pre-defined ontology is then translated into two needed lattices allowing to organize all the needed concepts and relations by using a hierarchical approach. The translation rules used are summarized in Figure 10 and Figure 11 and Figure 12 show a simplified version of each of these lattices.

| Object Diagram (UML) | | Lattice |
|---|---|---|
| Class | ↔ | Concept |
| Inheritance | ↔ | Concept hierarchy |
| Encapsulation | ↔ | Nested concepts |
| Method | ↔ | Relation |
| Relation | ↔ | Relation |
| Attribute | ↔ | Relation |

**Figure 10 :** Object Diagram / Lattice translation rules

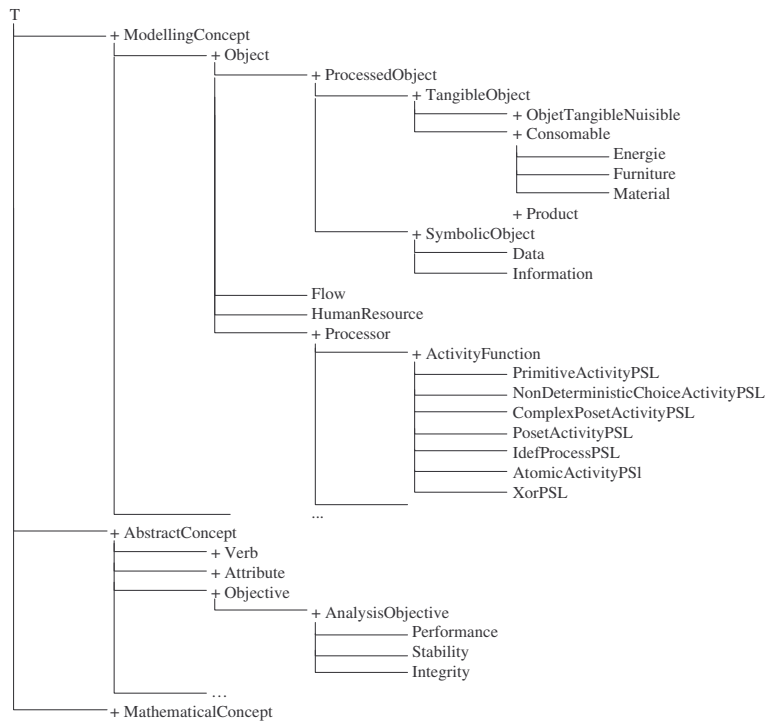

**Figure 11:** Extract from Concepts lattice

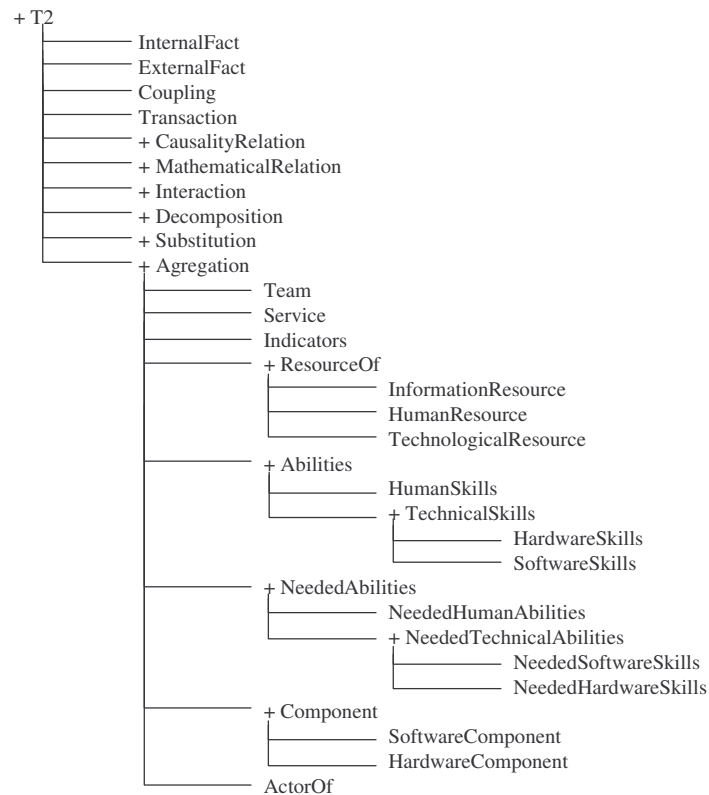

**Figure 12 :** Extract from Relations lattice

The Figure 13 illustrates how a given model is then translated by using these two lattices into a conceptual graph. Due to the paper size limitations, the translation rules are not presented here. The same rules are then used to translate each of the graph properties into several little conceptual graphs allowing the user to use now the formal mechanisms associated to the conceptual graphs in order to prove each of these property.

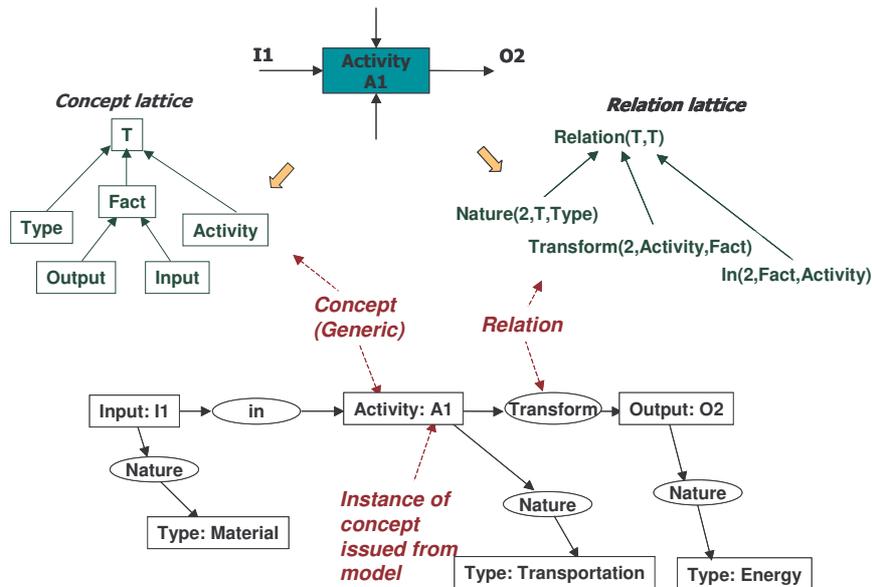

**Figure 13 :** Use of lattices for model translation into a conceptual graph

### 7.2. Reasoning concepts and mechanisms : First order logic isomorphism and canonical formation rules

(Sowa 1984) has established the existence of an isomorphism between the Conceptual Graphs and first order Logic. For that, an operator Φ has been defined in order to translate conceptual graphs into a set of first-order formulas manipulating predicate and logical relations:

- Each *conceptual relation* between two concepts is translated into an n-ary predicate

- Each *concepts* is transformed into an unary predicate

- Each *individual marker* becomes a constant

- Each *generic marker* becomes a quantified variable.

Thus, this operator provides a formal semantics to the conceptual graphs. In the following example, the sentence "James, the employee drills a part" which is represented by the conceptual graph G will have, as an equivalent logical formula Φ(G) :

$$G : [Employee : James] \rightarrow (agent) \rightarrow [Machine : drill] \leftarrow (object) \leftarrow [Part : *]$$

$$\Phi(G) : \exists x, (Employee(James) \wedge Machine(drill) \wedge Part(x) \wedge Agent (James, drill) \wedge Object(x, drill))$$

In addition, Sowa defined four elementary operations on the conceptual graphs, called *canonical formation rules*, which allows to handle them easily and to derive canonically from other graphs: *copy, restriction, simplification* and *joint*. These basic operations allows to handle and to transform graphs containing one information into other graphs that may contain a new unexpected information or which appeared not very interesting to the modeler: a new property can then emerge from the graph of existing properties and be thus proposed to this modeler.

### 7.3. Reasoning concepts and mechanisms : Projection

The handling operation called *projection* is the basic operation of a reasoning process in conceptual graphs. Projection corresponds to a graph morphism. The projection search of a graph G in a graph H can be seen as the inclusion search of the information represented by G in H. This leads to calculate a specialization between two graphs.

### 7.4. Reasoning concepts and mechanisms : Graph Rules

The conceptual graph *rules* (Salvat 1996) permit to represent knowledge in the form of inference rules of kind "If information H is in a graph, information C can be added to this graph". H and C are expressed as conceptual graphs related by co-reference links between some concept nodes. In other words a rule is composed of an hypothesis and a conclusion, and is used in the following classical way: given a simple graph, if the hypothesis of the rule projects to the graph, then the information contained into the conclusion is added to the graph. In this way, we define static rules to express some immutable domain laws and dynamic rules such as shown in Figure 14 to model the world evolutions with the change conditions of states.

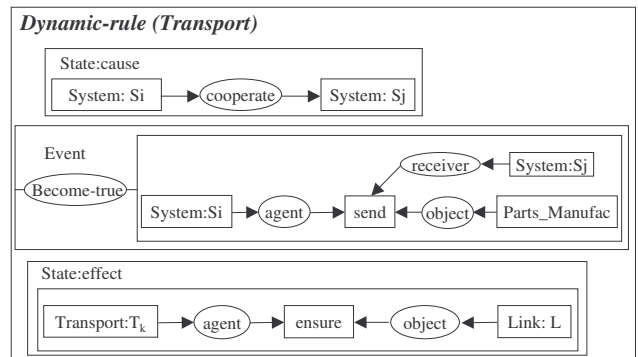

**Figure 14 :** Example of dynamic rule

## 7.5. Reasoning concepts and mechanisms : Constraints

A *constraint* defines conditions for simple graph to be valid. It is composed of a *condition* part and a *mandatory* part. The condition must be a simple graph. In particular, a condition can be an *empty graph*. Roughly said, a graph satisfies a constraint if for every projection of its condition part, its mandatory part also projects to the graph. We consider positive and negative constraints. A positive constraint expresses a property such as '*if information A is present, then information B must also be present*'. A negative constraint expresses a property such as '*if information A is present, then information B must be absent*'.

Hence, constraints are used to check validity of world. For example, let *G* be a simple graph. The graph *G* satisfies a positive constraint *Pc* if every projection of the condition of *Pc* into G can be extended to a projection of *Pc* as a whole. The graph *G* satisfies a negative constraint *Nc* if no projection of the condition of *Nc* into G can be extended to a projection of *Nc* as a whole. It allows to verify the graph G validity that is to say all the properties modeled as constraints are satisfied.

## 7.6. Application example

This way, conceptual graphs provides inference mechanisms for proving properties by using projection, rules and constraints. These demonstrative abilities make validation knowledge possible. For example, we consider an enterprise composed of some departments and of several workshops. The goal is to verify the property $P_1$ :

$P_1$  "*a department and a workshop have one person in common at the most* ".

For doing this, it is possible to use the negative constraint *Nc* (see) and the rule $R_1$ presented below:

*Nc*  "*incompatibility of membership relation with non-membership relation*"

$R_1$  "*if two persons x and y working in a workshop C are the members of a same department D, then all the persons of the workshop C are members of this department D* ".

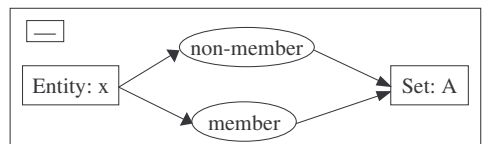

**Figure15 :** The negative constraint Nc

The proof by reductio ad absurdum of $P_1$ (by stating a proposition and then showing the resulting contradiction, thus demonstrating the proposition to be false) is as follows : supposing that in the company, there are two different persons *x* and *y* which both are the members of a department D and a workshop C, with C is not a subset of D. The previous rule $R_1$ tells us that all the persons of the workshop C are members of this department D, therefore C is a subset of D. This result belies our starting hypothesis, so it all goes to prove property $P_1$. This proof mechanism is formalized in conceptual graphs as follows:

- The starting hypothesis of proof is represented by the conceptual graph *Gh* in Figure 16.

- The result of the application of rule $R_1$ to graph *Gh* is represented by the conceptual graph *Gc* in Figure 17.

- The condition of negative constraint Nc is the empty graph, which can be projected into Gc. And this constraint is violated since there exists a projection of Nc as a whole into Gc. It's a contradiction in terms, so the proof of property $P_1$ is established.

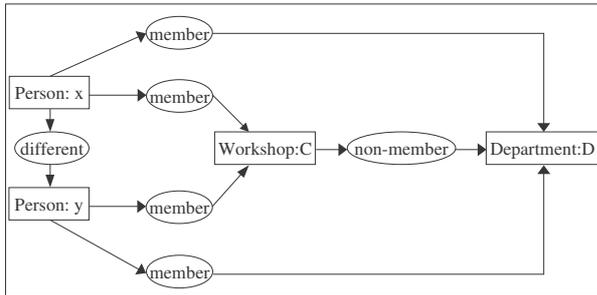 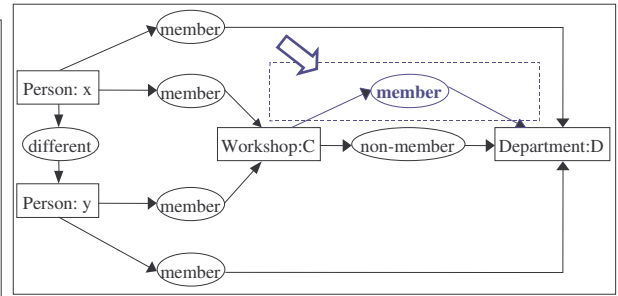

**Figure 16 :** Graph $G_h$ of the starting hypothesis to proof

**Figure 17 :** Graph $G_c$ resulting from the application of rule $R_1$ to graph $G_h$ reveals one contradiction

## 8. Conclusion and Perspectives

The property is an opportunity allowing an user to complete knowledge about a complex system and to formally investigate about it by using some formal mechanisms, concepts, ontology and tools supporting Verification and Validation step.

Evolution perspectives of this work are numerous but we will focus on the two following axes. First the proposed approach stays for the moment a little bit disturbing for an engineer or a consultant in charge of several processes. It is then necessary to develop some supporting and automated tools. Second it is necessary to integrate this approach in a global 'Enterprise Modeling and Analysis' method in which it will be possible to improve the V&V abilities of the main modeling approaches used in industry.